# On the Frequency-magnitude Law for Fractal Seismicity

## G. Molchan and T. Kronrod

*International Institute of Earthquake Prediction Theory and Mathematical Geophysics, Russian Academy of Sciences, Warshavskoye sh. 79, kor.2 Moscow, 113556, Russia.*
*The Abdus Salam International Centre for Theoretical Physics, Trieste, Italy*

E-mail: molchan@mitp.ru, kronrod@mitp.ru

*Abstract.* Scaling analysis of seismicity in the space-time-magnitude domain very often starts from the relation $\boldsymbol{l}(m,L) = a_L 10^{-bm} L^c$ for the rate of seismic events of magnitude $M > m$ in an area of size $L$. There are some evidences in favor of multifractal property of seismic process. In this case the choice of the scale exponent '$c$' is not unique. It is shown how different '$c$'s are related to different types of spatial averaging applied to $\boldsymbol{l}(m,L)$ and what are the '$c$'s for which the distributions of $a_L$ best agree for small $L$. Theoretical analysis is supplemented with an analysis of California data for which the above issues were recently discussed on an empirical level.

### 1. Introduction

The rate of seismic events of magnitude $M > m$ occurring in a cell of size $L \times L$ denoted $\boldsymbol{l}(m, L)$ is *a priori* scaled as follows:

$$\boldsymbol{l}(m,L) = a 10^{-bm} L^c. \qquad (1)$$

The magnitude-dependent exponential factor stems from the Gutenberg-Richter relation, while the power law factor, which is a function of area size, expresses the fractality of epicenters for a noninteger '$c$'. Relation (1) is given the meaning of a seismicity law in [1, 2] and a method is proposed for estimating its parameters ($a$, $b$, $c$). Viewed as such, relation (1) needs specification, since a law must characterize a mean or "typical" earthquake-generating area in a region of interest. Below we show that different specifications may lead to different values of '$c$'.

Our analysis of (1) was occasioned by the circumstance that the estimation procedure proposed for '$c$' in [1, 2] leads to a correlation dimension $d_2$, while the motivation of scaling (1) is based on the capacity (box) dimension $d_0$. A similar difficulty with the choice of '$c$' was encountered when scaling the time interval between two consecutive events in California: Bak et al. [3] used the estimate $c = d_2$, while subsequent works dealing with the topic made use of $c = d_0$ (see Corral [4]). It has turned out that the estimates $d_0$ and $d_2$ are not identical. For instance, the same California catalog gave $d_2 = 1.2$ [2, 5] and $d_0 = 1.6$ [4].

The dimensions $d_0$ and $d_2$ belong to the one-parameter family of the so-called Grassberger-Procaccia dimensions [6], $d_p$. These dimensions are strictly decreasing, if the measure of the rate of $M > m$ events denoted $\boldsymbol{l}(dg \mid m)$, i.e., the mean number of events per unit time in an area $dg$, is a multifractal. Since the above estimates $d_0$ and $d_2$ are not identical, we will consider relation (1) in terms of a multifractal hypothesis for the measure $\boldsymbol{l}(dg \mid m)$.

More specifically, we are going to find suitable exponents '$c$' for different types of averaging applied to the quantities $\boldsymbol{l}(m,L)$ and for histogram of these quantities. The theoretical analysis of the population $\boldsymbol{l}(m,L)$ will be illustrated by consideration of California seismicity.

## 2. Scalings for multifractal seismicity

### 2.1. The measure $l(dg \mid m)$ as a multifractal

We use a rectangular grid to partition a region $G$ into $L \times L$ cells. Let $l_G(m)$ be the rate of $M > m$ events in $G$, and let $l_i(m, L)$ be that for the $i$-th $L \times L$ cell. The number of cells having positive $l_i$ is denoted $n(L)$. If the relation

$$\log n(L) = -d_0 \log L \,(1+o(1)), \quad L \to 0, \quad 0 < d_0 < 2, \tag{2}$$

holds, then it is said that the support of the measure $l(dg \mid m)$ is fractal and has a box dimension $d_0$. When $l(dg \mid m)$ is multifractal, the support is stratified, roughly speaking, into a sum of fractal subsets $S_a$ having the dimensions $f(a) \in (\underline{d}, \overline{d})$. The points in $S_a$ are centers of concentration for epicenters, so that one has

$$\log l(m, L) = a \log L \,(1+o(1)). \tag{3}$$

in a sequence of $L \times L$ areas (as $L \to 0$) that contain a concentration point. Relation (3) describes a type of spatial concentration of events or a type of singularity for $l(dg \mid m)$. Accordingly, $f(a)$ describes the box/Hausdorff dimension of centers having the singularity type $a$. Pairs $(a, f(a))$ form a multifractal spectrum of the measure $l(dg \mid m)$. Information on the multifractal behavior of $l(dg \mid m)$ can be gathered from the Renyi function:

$$R_L(p) = \sum_{l_i > 0} (l_i(m,L)/l_G)^p, \quad |p| < \infty, \tag{4}$$

which admits of the asymptotic expression

$$\log R_L(p) = t(p) \log L \,(1+o(1)), \quad L \to 0, \tag{5}$$

where the scaling exponent $t(p)$ being closely related to $f(a)$ by the Legendre transform:

$$t(p) = \min_a (pa - f(a)). \tag{6}$$

When $p = 0$, relation (5) becomes (2), hence $t(0) = -d_0$. In the case of a monofractal measure when the interval $[\underline{d}, \overline{d}]$ degenerates into the point $d_0$, the function $t(p) = d_0(p-1)$ is linear. In the general case $t(p)$ is convex upwards, and $t(1) = 0$. If $t(p)$ is strictly convex and smooth, the range of values of derivative $t(p)$ defines the interval of possible $a$ singularities in (3), while the Legendre transform of $t(p)$: $\min_p (pa - t(p)) = f(a)$ describes the dimensions of these singularities. The above statements constitute multifractal formalism [7] whose mathematical content is more profound and has limitations of its own.

The quantities $d_p = t(p)/(p-1)$ are known as generalized Grassberger-Procaccia dimensions. From the relation $t(1) = 0$ and the mean value theorem one has

$$d_p = \frac{t(p) - t(1)}{p - 1} = t(p^*), \tag{7}$$

where $p^*$ is a point between 1 and $p$. Consequently, in the case of smooth and strictly convex $t(p)$, $d_p$ describes a type of singularities or a "local dimension" of $l(dg \mid m)$.

### 2.2. Scaling of the averaged $l_i(m, L)$

Let us characterize the rate of $M > m$ events in an $L \times L$ cell of the region $G$ by averaging the $l_i(m, L)$ over all cells with some weights. The choice of weights depends on the purpose for which we wish to use the mean. One sufficiently flexible and natural family to use is the one-parameter family of weights

$$m_i^{(p)} = k_p \mathbf{1}_i^p, \quad |p| < \infty, \quad \mathbf{1}_i > 0,$$

where $k_p$ is a normalizing constant such as to make $\sum m_i^{(p)} = 1$. By (4) one has

$$1/k_p = R_L(p) \mathbf{1}_G^p.$$

When $p = 0$, one has ordinary averaging of $\mathbf{1}_i(m,L)$ with $\mathbf{1}_i > 0$ while when $p \gg 1$, the mean will characterize the most active cells, because $\sum \mathbf{1}_i m_i^{(p)} \to \max_i \mathbf{1}_i$, as $p \to \infty$.

Consider the mean $<.>_p$ with weights $m_i^{(p)}$. In that case

$$<\mathbf{1}_i(m,L)>_p = \sum_{\mathbf{1}_i > 0} \mathbf{1}_i m_i^{(p)} = \mathbf{1}_G(m) R_L(p+1)/R_L(p).$$

If (5) holds, then

$$\log <\mathbf{1}_i(m,L)>_p = [\mathbf{t}(p+1) - \mathbf{t}(p)] \log L (1 + o(1)) + \log \mathbf{1}_G(m)$$

or

$$<\mathbf{1}_i(m,L)>_p \sim \mathbf{1}_G(m) L^{c_p}, \qquad (8)$$

where $c_p$ has the nontrivial form

$$c_p = \mathbf{t}(p+1) - \mathbf{t}(p) = p d_{p+1} - (p-1) d_p. \qquad (9)$$

When the region of interest is large, $\mathbf{1}_G(m)$ is satisfactorily described by the Gutenberg-Richter frequency-magnitude relation $\mathbf{1}_G(m) = a 10^{-bm}$, so that (8, 9) constitute a refined variant of (1) for the case of the multifractal measure $\mathbf{1}(dg \mid m)$.

One is mostly interested in the averaging with $p = 0$ and $p = 1$. In that case

$$c_p = \begin{cases} d_0 \\ d_2 \end{cases} \text{is} \quad \begin{array}{l} \text{box dimention,} \quad p = 0 \\ \text{correlation dimention,} \quad p = 1. \end{array}$$

Thus, the box dimension is relevant to ordinary averaging $<\mathbf{1}_i>_0$, while the correlation dimension $c = d_2$ is relevant to the averaging that is proportional to the rate of events in each $L \times L$ cell. The weights $\{m_i^{(p)}\}$ can be interpreted as the probability distribution $P_L^{(p)}$ to have in mind when making the choice of an $L \times L$ cell. In that case (8) describes the rate of $M > m$ events in $P_L^{(p)}$ – random $L \times L$ cell in the region $G$. Similarly to (7), one infers that

$$c_p = \mathbf{t}(p+1) - \mathbf{t}(p) = \mathbf{t}(p + \mathbf{d}^*), \quad 0 \leq \mathbf{d}^* \leq 1,$$

that is, $c_p$ can correspond to some local dimension of $\mathbf{1}(dg \mid m)$. *The interpretation of 'c' in terms of box dimension $d_0$ is possible either for the monofractal measure $\mathbf{1}(dg \mid m)$ or for the equip probable choice of the earthquake-generating cell.*

### 2.3. Scaling the distribution of $\mathbf{1}(m,L)$

Consider the population of normalized $\mathbf{1}(m,L)$: $\mathbf{x}_L = \{\mathbf{1}_i(m,L)/10^{-bm} L^c\}$, related to the subdivision of region $G$ into $L \times L$ cells. The distribution of these quantities provides another statistical description of $M > m$ seismicity rate in an $L \times L$ area in $G$. Corral [4] found that the distribution of $\mathbf{x}_L$ for California is virtually independent of the parameter $L$ in the range 10–120 km for $m = 2$ and 3. The $b$-value in the Gutenberg-Richter relation was taken 0.95, while the scale exponent $c = d_0 = 1.6$. It is also asserted in [4] that the distribution of $\mathbf{x}_L$ is only weakly dependent on the choice of the time interval $\Delta T$ in the range of 1 day to 9 years. The statement about $\Delta T$ calls for some specification in order to be reproducible. Nevertheless, the following question arises for a multifractal measure $\mathbf{1}(dg \mid m)$: for what values of 'c' does the distribution of $\boldsymbol{x}_L$ have a limit as $L \to 0$? With these 'c' one is entitled to expect that the distributions of $\boldsymbol{x}_L$ are similar for small $L$.

Similarly to Section 2.2, we will extend the problem by using the weights $m_i^{(p)} = k_p \boldsymbol{l}_i^p$ as a probability measure $\mathrm{P}_L^{(p)}$ for $\boldsymbol{x}_L$. When $p = 0$ therefore, we arrive at the distribution of $\boldsymbol{x}_L$ which was considered in [4].

We begin by considering an example. Suppose the measure $\boldsymbol{l}(dg \mid m)$ has density $f(g)$; the distribution of $\boldsymbol{x}_L$ then converges to a distribution of the form

$$F(x) = mes\{g : 0 < f(g) < 10^{-bm} x\} / mes\{g : f(g) > 0\},$$

as $L \to 0$ in the case $c = d_0 = 2$. The limit is independent of the choice of the subdivision grid for $G$. Here, mes(A) is the area of region A.

The class of multifractal measures is very broad, while the measures themselves may have very complicated structure. For this reason we shall provide standard heuristic arguments to find a suitable $c = c^{(p)}$ for a given p, so that one can expect a nontrivial limiting distribution for $(\boldsymbol{x}_L, \mathrm{P}_L^{(p)})$.

Denote the multifractal spectrum of $\boldsymbol{l}(dg \mid m)$ by $f(\boldsymbol{a})$. The number of $L \times L$ cells of type $\boldsymbol{a}$, i.e., such that $\boldsymbol{l}_i(m,L) \sim L^{\boldsymbol{a}}$, is increasing like $L^{-f(\boldsymbol{a})}$. Consequently, $\boldsymbol{l}_i(m,L)/L^c$ is bounded away from 0 and $\infty$ as $L \to 0$, if the i-*th* cell belongs to type $\boldsymbol{a} = c$. The probability or weight of cells of type $\boldsymbol{a}$ is of the order

$$L^{-f(\boldsymbol{a})} m_i^{(p)} = L^{-f(\boldsymbol{a})} \boldsymbol{l}_i^p(L)/R_L(p) \sim L^{-f(\boldsymbol{a}) + p\boldsymbol{a}} / L^{\boldsymbol{t}(p)},$$

where $R_L(p)$ is given by (4), while $\boldsymbol{t}(p)$ is $\boldsymbol{t}(p) = \min_{\boldsymbol{a}}(p\boldsymbol{a} - f(\boldsymbol{a}))$ (see (6)). The resulting probability is bounded away from 0 as $L \to 0$, only if $\boldsymbol{t}(p) = p\boldsymbol{a} - f(\boldsymbol{a})$. Consequently, the desired $c = c^{(p)}$ is such that $p\boldsymbol{a} - f(\boldsymbol{a})$ reaches its minimum when $\boldsymbol{a} = c$; in short,

$$c^{(p)} = \arg \min_{\boldsymbol{a}} (p\boldsymbol{a} - f(\boldsymbol{a})).$$

In particular, when $p = 0$, the desired $c^{(0)}$ is the point of maximum for $f(\boldsymbol{a})$, i.e.,

$$c^{(0)} \text{ is the root of the equation } f(\boldsymbol{a}) = d_0. \tag{10}$$

If spectrum $f(\boldsymbol{a})$ is a strictly convex function, it can be described parametrically in terms of $\boldsymbol{t}(p)$: $\boldsymbol{a} = \boldsymbol{t}(p)$, $f(\boldsymbol{a}) = p\boldsymbol{a} - \boldsymbol{t}(p)$.
Hence

$$c^{(p)} = \boldsymbol{t}(p). \tag{11}$$

In the example considered above, spectrum $f(\boldsymbol{a})$ consists of the single point $(\boldsymbol{a}, f(\boldsymbol{a})) = (2,2)$. Consequently, $c^{(p)} = d_0 = 2$. Now consider a more complex example, namely, a measure with density on the cell $[0, 1]^2$ and in the interval $[1, 2]$. This is a "fractal" mixture with two points in the spectrum $(\boldsymbol{a}, f(\boldsymbol{a}))$: (2, 2) and (1, 1). When $0 \le p < 1$, we get $c^{(p)} = d_0 = 2$, and $c^{(p)} = 1$ when $p > 1$. Relation (11) does not work at $p = 1$, because $\boldsymbol{t}(p)$ is no longer smooth: $\boldsymbol{t}(1-0) = 2 \ne \boldsymbol{t}(1+0) = 1$.

In the examples considered here, equation (10) has the solution $c^{(0)} = d_0$. In the general case one can only assert that $c^{(0)} \ge c_0 = d_0$. This can be seen as follows. The function $\boldsymbol{t}(p)$ is convex upwards. Therefore, $\boldsymbol{t}(p)$, $0 < p < 1$ lies above the chord that connects the points $(0, \boldsymbol{t}(0))$ and $(1, \boldsymbol{t}(1))$, i.e.,

$$\boldsymbol{t}(p) \ge \boldsymbol{t}(1)p + (1-p)\boldsymbol{t}(0) = (1-p)\boldsymbol{t}(0), \quad 0 \le p \le 1$$

and so $c^{(0)} = \bm{t}(+0) \geq -\bm{t}(0) = d_0$.

It is for the same reason that $\bm{t}(p)$ lies below the tangent at any point $p$, i.e.,

$$\bm{t}(p) \leq \bm{t}(0) + \bm{t}(+0) p = -d_0 + c^{(0)} p.$$

Consequently, *if $f(d_0) = d_0$, then $\bm{t}(p) = d_0(p-1)$ for all $0 \leq p \leq 1$.*

This simple remark can conveniently be used to verify the equality $c^{(0)} = d_0$, since $\bm{t}(p)$ is much more accurately calculated for $p > 0$ than is the case for $\bm{t}(0) = -d_0$ and $\bm{t}(0) = c^{(0)}$.

To sum up, we have arrived at two inquisitive scaling relations:

$$\langle \bm{l}_i(m,L) \rangle_0 \sim L^{c_0}, \quad L \to 0$$

and

$$\text{the histogram of } \{\bm{l}_i(m,L)\} \sim L^{c^{(0)}}, \quad L \to 0 \tag{12}$$

with (generally speaking) different exponents '$c$': $c^{(0)} \geq c_0 = d_0$.

The paradox is easily resolved. In the second of these relations the choice of $c = c^{(0)}$ ensures the convergence of the distribution of $\bm{x}_L$ as $L \to 0$; at the same time, $\bm{l}_i(m,L)$ of type $\bm{a} < \bar{d} = c^{(0)}$ asymptotically give zero contribution in the limit. For other $c \neq c^{(0)}$ the limiting distribution of $\bm{x}_L$ degenerates, being concentrated at 0 and $\infty$. The contribution of all $\lambda_i(m,L)$ of type $\alpha = c^{(0)} = \bar{d}$ into the average $<\bullet>_0$ is of order $L^{c^{(0)}}$. It is for this reason that $L^{c_0} \geq L^{c^{(0)}}$ as $L \to 0$.

In practical terms, the difference between $c_0$ and $c^{(0)}$ may be small. For, expressing them through $\bm{t}(p)$ in the general case where the $\bm{l}_i(m,L)$ are used with the weights $m_i^{(p)} = k_p \bm{l}_i^p$, one has from (8, 9):

$$\langle \bm{l}_i(m,L) \rangle_p \sim L^{c_p}, \quad c_p = \bm{t}(p+1) - \bm{t}(p) = \bm{t}(p + \bm{g}_p), \quad 0 < \bm{d}_p < 1.$$

At the same time, the optimal scale exponent for the distribution $\{\bm{l}_i(m,L), \mathrm{P}_L^{(p)}\}$ is $c^{(p)} = \bm{t}(p)$, see (11). Hence

$$c^{(p)} = \bm{t}(p) \geq \bm{t}(p + \bm{q}_p) = c_p \text{ for all } p \geq 0. \tag{13}$$

For California seismicity with $m \geq 2$, Corral [4] found that the distributions of $\bm{x}_L$ are well consistent in a broad range of $L$ using $c = d_0 = 1.6$. That may mean that $c^{(0)} = d_0 = 1.6$. We shall try to verify the above conclusion in the section to follow.

### 3. California Seismicity

We used the catalog of $m \geq 2$ California events for the period 1984-2003 [8] in the rectangle $G = (30°\text{N}, 40°\text{N}) \times (113°\text{W}, 123°\text{W})$. Estimation of the '$b$'-value in the Gutenberg-Richter relation does not cause any difficulties, and we adopted $b = 0.95$ for $G$. The estimation of $d_0$ is unstable, so the estimation procedure is described below. As pointed out above, the fractal dimension 1.2 is used for '$c$' in [3] for scaling of interoccurrence time between earthquakes, while $c = d_0 = 1.6$ is assumed in the sequel [4] without indicating the estimation method.

*The box dimension $d_0$* is given by (2). The principal difficulty in estimation of $d_0$ for point sets consists in their finiteness. The number of cells is increasing like $L^{-2}$ as $L \to 0$. For this reason the number of cells $n(L)$ that cover our set rapidly saturates, providing the false (even though formally correct) estimate $d_0 = 0$. The epicenters of seismic events are special in the sense that they make a random set. Owing to purely statistical factors, some of the seismogenic cells for small $L$ are empty because of the low rate $\boldsymbol{l}(m,L)$. The situation becomes critical, when the empty cells $n_0$ make an appreciable part of $n(L)$ ($n_0/n(L) > \boldsymbol{e}$, say). In that case the loss of $n_0$ cells will noticeably affect the estimated slope of $(\lg L^{-1}, \lg n(L))$. We try to find the critical scale $L^*$ by computing the statistic $n(L,k)$ with $k = 0, 1,...$ which gives the number of $L \times L$ cells that have numbers of events $>k$. In this notation $n(L, 0)=n(L)$. The quantity $n_1 = n(L) - n(L,1)$ will give the number of cells having the number of events equal to 1. The statistical nature of numbers of events 1 or 0 in a seismogenic cell is one and the same: a low rate of events, more specifically, $\boldsymbol{l}(m,L)\Delta T \leq 1/2$. It would therefore be natural to expect that $n_1$ and $n_0$ have the same order of magnitude. In that case however the requirement $n_0/n(L^*) = \boldsymbol{e}$ can be replaced with

$$\frac{n(L^*) - n(L^*,1)}{n(L^*)} = \boldsymbol{e}, \tag{14}$$

which specifies the critical value of $L$.

Leaving aside for the moment the stochastic nature of epicenters, requirement (14) means that the desired estimate of $d_0$ should be little sensitive to cells with low numbers of events. (This principle is used later on to estimate other dimensions). We use $\boldsymbol{e} = 10\%$ in our calculations. If close-lying pairs of events are highly probable for a random set, then it is natural to use $n(L^*, 2)$ instead of $n(L^*, 1)$ in (14).

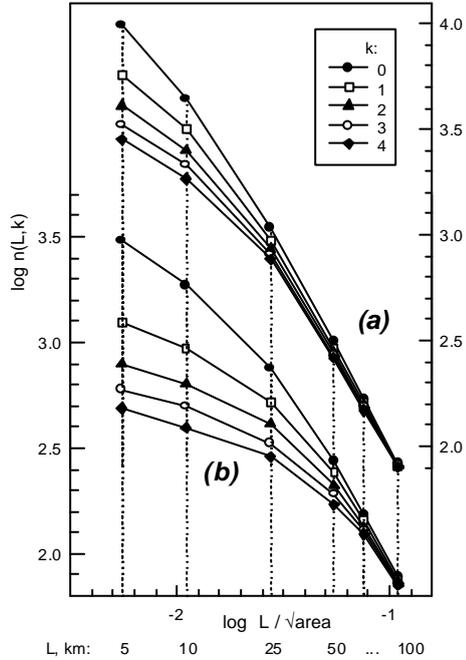

Figure 1. Data for estimating the box dimension of earthquake epicenters with $m \geq 2$ and $m \geq 3$ for California.

The vertical axis shows the number of $L \times L$ cells with the number of events $>k$, $k = 0, 1, 2, 3, 4$ and magnitude $m \geq 2$ (a) and $m \geq 3$ (b). The total number of events: 116710 (a) and 11783 (b). Vertical axis: for (a) on the right and for (b) on the left.

Figure 1 shows curves of $n(L^*,k)$ for $m \geq 2$ and $m \geq 3$ events in California. It appears from these plots that the critical scale is $L^* = 25$ km for $m \geq 2$ and $L^* = 50$ km for $m \geq 3$. Estimation of $d_0$ from $n(L)$ in the interval $(L^*, 100$ km$)$ gives $d_0 = 1.9$. Various translations and rotations of the subdivision grid for $G$ leaves the estimate of $d_0$ in the range 1.8-1.9.

*The distributions of $\boldsymbol{x}_L$.* Several estimates of the fractal dimension of epicenters are available for scaling the distribution of $\boldsymbol{x}_L$: $d_0 = 1.8-1.9$ (as found above), $d_0 = 1.6$ [4], and $d_2 = 1.2$ [2, 5]. The dimensions $d_0$ and $d_2$ were both used for seismicity scaling as an anonymous fractal dimension (see [2], [3]). The situation becomes more complicated, since the recent work [9] gives 1.5−1.7 as estimates of the correlation dimension for mainshock hypocentres. When converted to the dimension of epicenters therefore, one should expect $d_2 \approx 0.5 - 0.7$. The question about the suitable scaling of $\boldsymbol{l}(m,L)$ remains therefore essentially unresolved.

Figure 2 shows histograms of

$$\lg \boldsymbol{x}_L = \left\{ \log \frac{n_i(m,L)/T}{10^{-bm}(L/L_0)^c} \right\}, \tag{15}$$

where $n_i(m,L) > 0$ is the number of events in the i-*th* $L \times L$ cell during the time $T = 20$ years, and $L_0^2 = 82645$ km$^2$ is the area of the region $G$. The parameters involved are $L = 10, 25, 50, 70$ and $100$ km and $c = 1.2$ (a), $1.6$ (b), $1.8$ (c), $2.0$ (d). The other 'c' parameters are omitted for reasons of space. The histograms of $\lg \mathbf{x}_L$ are shown for $m \geq 2$ only, the data for $m \geq 3$ being scanty. We consider the population $\lg \mathbf{x}_L$ instead of $\mathbf{x}_L$, because the scaling of $\mathbf{l}_i(m,L)$ is more meaningful when viewed in a log scale. Theoretically speaking, the densities of $\mathbf{x}_L$ and $\lg \mathbf{x}_L$ differ by a linear function having a slope of 1 in a *log-log* plot.

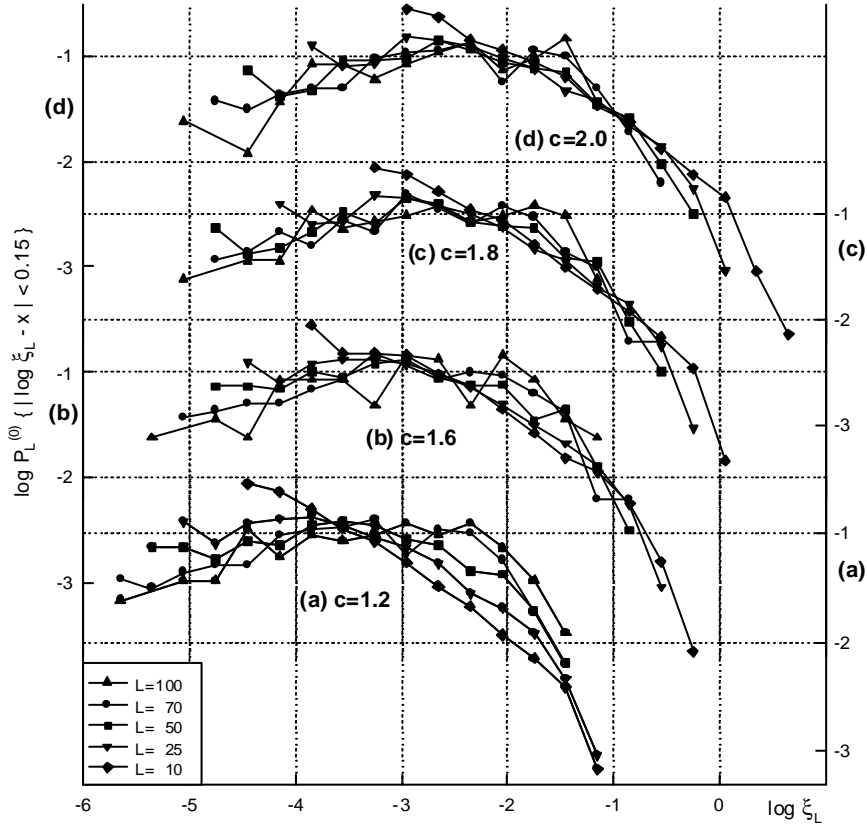

Figure 2. Histograms of $\lg \mathbf{x}_L$ with parameters c=1.2 (a), 1.6 (b), 1.8 (c) and 2.0 (d) for cells of size *L*: 10, 25, 50, 70, 100 km

The series of curves (b), (c) and (d) have been shifted vertically relative to (a) by the amounts 1.5, 3.0, and 4.5 log units, respectively. For convenience each curve has a vertical scale of its own attached: (b), (d) on the left, (a), (c) on the right.

As appears from Fig. 2, the histograms of $\lg \mathbf{x}_L$ are fairly well consistent for different *L*. The agreement seems to be the best for $c = 1.8–2.0$, i.e., $1.6 < c^{(0)} \leq 2$.

We are going to show that $c^{(0)} > d_0$. To do this, we find the generalized dimensions $d_p = \mathbf{t}(p)/(p-1)$, $0 < p < 1$. As mentioned above, if $c^{(0)} = d_0$, then $d_p$ is constant in (0, 1). Figure 3 provides estimation of $\mathbf{t}(p)$ for $p = 0.25, 0.5, 0.75$, showing plots of $R_L(p)$ (see (4)) which were, as in the case of $d_0$, computed in $L \times L$ cells that contain more than $k$ events, $k$ taking on the values 0, 1, 2, 3, 4. (These modifications of the Renyi function are denoted $R_L(p,k)$). Our estimation of $\mathbf{t}(p)$ is based on the slope of $(\log L, \log R_L(p,0))$ in the scale range $L = 20–100$ km where the cells with a single event do not affect the results. Figure 3 gave the following table:

| $p$   | 0.25 | 0.50 | 0.75 |
|-------|------|------|------|
| $d_p$ | 1.71 | 1.64 | 1.48 |

from which it appears that $d_p$ is not constant in [0, 1].

It follows that $2 \geq c^{(0)} > c_0 = d_0 = 1.8 - 1.9$; the scale exponent $c = 1.8–2.0$ is equally well suitable for scaling of both the mean $\langle \mathbf{l}_i(m,L) \rangle_0$ and the distribution of $\{\mathbf{l}_i(m,L)\}$ for $m = 2$. Because '$c$' is close to 2, the role of fractality in scaling $\mathbf{l}(m,L) > 0$ is unessential.

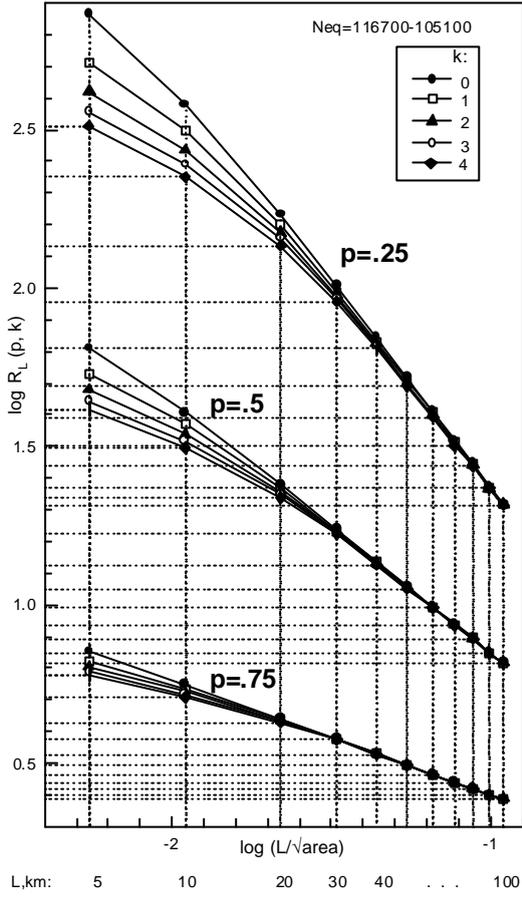

Figure 3. Data for estimating $\tau(p)$, $p=0.25, 0.50, 0.75$ for $m \geq 2$ events in California

The vertical axis shows modified Renyi functions $R_L(p,k)$ (see (4)) based on data in $L \times L$ cells having the number of events $> k = 0, 1, 2, 3, 4$.

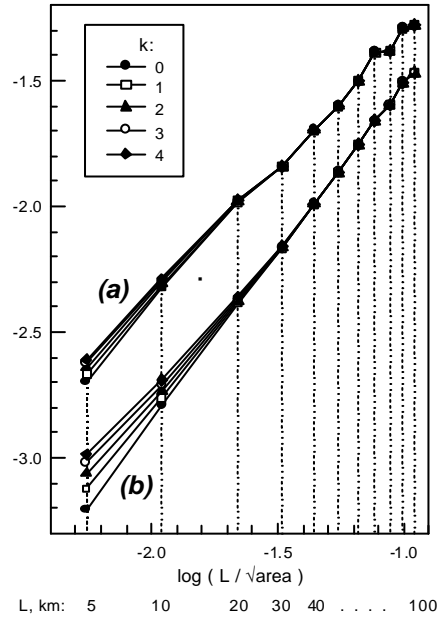

Figure 4. Data for estimating the correlation dimension $c_1 = d_2$ (a) and $\dot{\tau}(1) = c^{(1)}$ (b)

Shown along the vertical axis are (a) $\log R_L(p,k)$ and (b) $\log \dot{R}_L(p,k)$, where $R_L(p,k)$ are modified Renyi functions based on $L \times L$ cells with numbers of events $> k = 0, 1, 2, 3, 4$.

The distributions of $\lg \mathbf{x}_L$ (see (15)) with the exponent $c = 1.6$ (Fig. 2b) are far from the perfect agreement at different scales reported in [4]. We therefore prefer the estimate $d_0 = 1.8$ for $m \geq 2$.

*The weighted scaling of $\mathbf{l}(m, L)$*. The foregoing analysis concerns the scaling of $\mathbf{l}(m, L) > 0$ in a random $L \times L$ cell irrespective of its contribution into the overall seismicity. Consider the scaling of $\mathbf{l}(m, L)$ for the case in which the i-*th* cell is sampled with a probability proportional to $\mathbf{l}_i(m, L)$. In that case the scale exponent for the mean $\langle \mathbf{l}_i(m, L) \rangle_1$ is identical with the correlation dimension, $c_1 = d_2$. The data for estimating $d_2$ can be seen in Fig. 4a. Since $d_2 = \boldsymbol{\tau}(2)$, the estimation procedure for $d_2$ is the same as in Fig. 3. Figure 4a corroborates the estimate $d_2 = 1.1-1.2$ [2, 5], well known for California. The optimal exponent '$c$' for scaling of the distribution of $\{\mathbf{l}_i(m, L)\}$ is $c^{(1)} = \boldsymbol{\tau}(1)$. It is found as the slope of $(\lg L / L_0, \lg \dot{R}_L(1))$ where $\dot{R}_L(p)$ is the derivative of the Renyi function with respect to $p$ (see Fig. 4b). Figure 4b also shows, for comparison purposes, the modified Renyi functions, i.e., $\dot{R}_L(p,k)$, $k = 0, 1, 2, 3, 4$. From Fig. 4b follows a reliable estimate of $c^{(1)}$: $c^{(1)} = 1.3 - 1.4 > c_1 = 1.1 - 1.2$.

The histograms of $\lg \mathbf{x}_L$ derived with the weights $w_i = k \mathbf{l}_i(m, L)$ are shown in Fig. 5 for a range of scale exponent, $c = 1.2$–$2.0$. The histograms look the least consistent at $c = 1.2$. When, on the other hand, one uses only the weightier points in the histogram, i.e., those with the mass $\geq 0.01$ (see the vertical axis), then the scatter in the distribution of $\lg \mathbf{x}_L$ is the least for $c \leq 1.6$. For this reason Fig. 5 provides an independent estimate of $c^{(1)}$ as the interval $1.3 < c^{(1)} < 1.7$ for the limiting distribution $\{\mathbf{l}_i(m, L), P_L^{(1)}\}$ with $L = 10$–$100$ km.

Consequently, the scaling of $I(m, L)$ turns out to be rather indeterminate, since $c_1 = 1.1 - 1.2$ and $c^{(1)} = 1.3 - 1.7$.

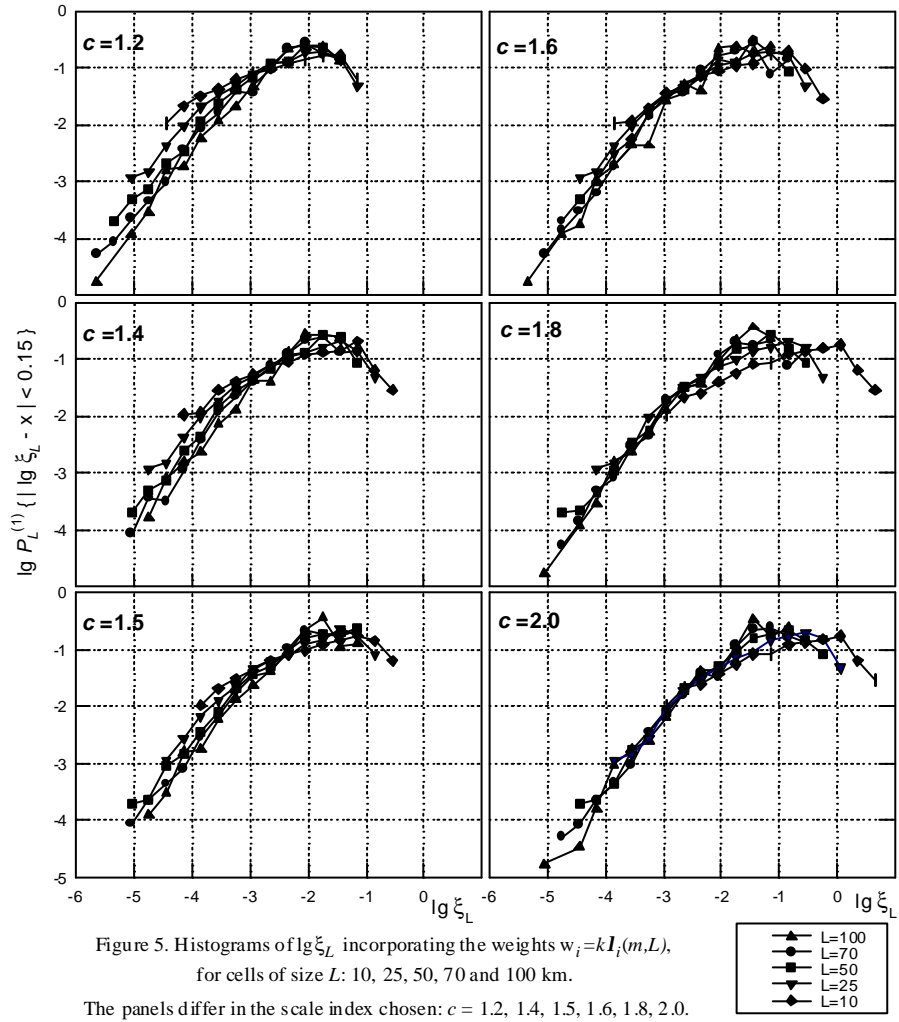

Figure 5. Histograms of $\lg \xi_L$ incorporating the weights $w_i = k I_i(m,L)$, for cells of size $L$: 10, 25, 50, 70 and 100 km.
The panels differ in the scale index chosen: $c$ = 1.2, 1.4, 1.5, 1.6, 1.8, 2.0.

### 4. Scaling and magnitude: discussion

In our analysis the cutoff magnitude $m$ is fixed, so that the question as to the relation between the scale exponent '$c$' and the distribution of $x_L$ with $m$ was not discussed. In this connection we wish to point out the following. Great earthquakes usually occur at intersections of lineaments of the highest rank [10], large ones on lineaments themselves, while smaller events are diffused over the entire seismogenic region concerned. In this respect one notes Fig. 6 showing larger Italian earthquakes. In contrast to the standard situation then fractal analysis is based on catalogs of small events for a short period of time, Figure 6 shows largest events from the catalogue [3] for a nearly 1000-year period, 1000 to 1980. Earthquake size is characterized (because of natural reasons) in terms of macroseismic intensity I: I > 7 (a), I > 8 (b), and I > 9 (c). Figure 6 clearly shows differences in seismicity generators: the largest events concentrate along a narrow belt (McKenzi boundary) of width 30-50 km, while smaller events make the boundary more diffuse, thus inflating $d_0$. It may therefore be conjectured that we have here a mixture of monofractals corresponding to different sets of magnitude, while the measure $I(dg \mid m)$ is function of $m$. The circumstance is commonly disregarded, so that relations like (1) are extrapolations from small $m$ to high magnitudes.

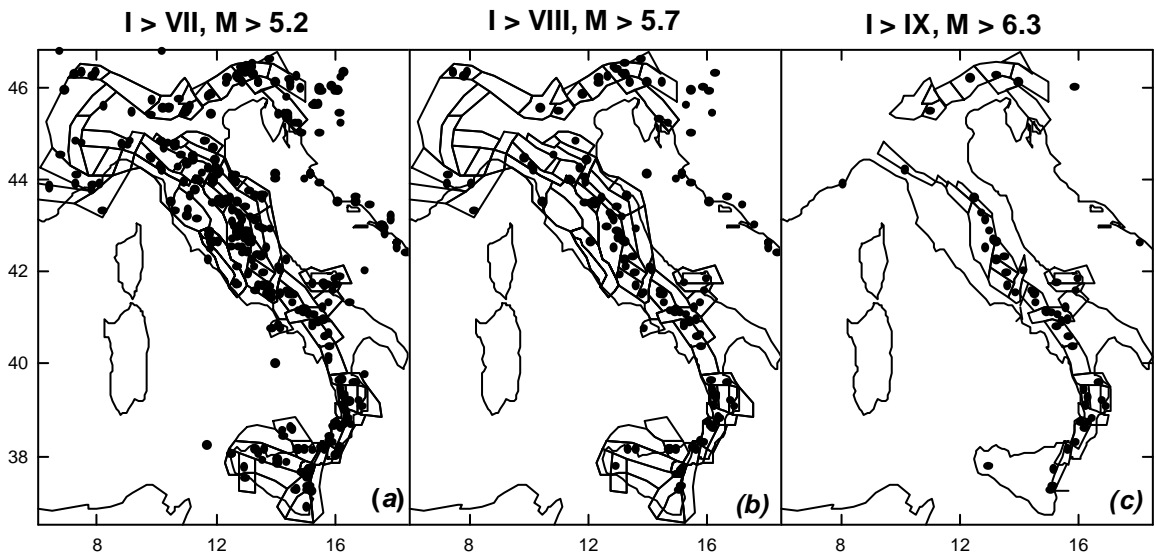

Figure 6. Large Italian earthquakes for the period 1000-1980 based on the Stucchi et al. (1993) catalog, and the earthquake-generating zon

When the frequency-magnitude relation $\boldsymbol{l}_G(m)$ in a region $G$ is described by the Gutenberg-Richter law: $a10^{-bm}, m \in \Delta M$, then also here, problems can arise with the uniformity of the parameter $b$ for all magnitudes. A typical limitation for the above description sounds as follows: the linear size of $m \in \Delta M$ events is much smaller than the linear size of region $G$ and the thickness of seismogenic layer [14]. Otherwise one can encounter phenomena like characteristic earthquakes which distort the straight line $\log \boldsymbol{l}_G(m)$ for large $m$.

### 5. Conclusion

We have ascribed a definite meaning to relation (1) which is frequently used in seismicity studies, namely, for unification of distributions of different statistics depending on scale and magnitude [3], in earthquake prediction [10, 11], and in aftershock identification [12]. When the seismicity field is multifractal, the choice of '$c$' in (1) is nonunique which is related to different interpretations of $\boldsymbol{l}(m, L)$ as the rate of $M > m$ seismic events in a "random" $L \times L$ cell of the region of study. We have shown using the California data with $m = 2, 3$ that the scale exponent '$c$' may vary in the range 1–2. In particular, $c = 1.8$–$2.0$ is suitable for scaling of both the ordinary mean and the distribution of $\boldsymbol{l}(m, L) > 0$ in $L \times L$ cells. (The value $c = 1.6$ is used in recent studies of California seismicity [2, 4, 11, 12].) But we can solve these scaling problems using weights proportional to $\boldsymbol{l}(m, L)$ in $L \times L$ cells. This practice is typical for statistical evaluation of performance of earthquake prediction algorithms (see [10]). Then one has $c = 1.1$–$1.2$ for the scaling of the weighted mean $\langle \boldsymbol{l}_i(m, L) \rangle_1$ and $c = 1.4$–$1.6$ to have the least scatter among the normalized distributions $\{\boldsymbol{l}_i(m, L), \mathrm{P}_L^{(1)}\}$ with the above weights.

This large indeterminacy in the choice of '$c$' is extremely inconvenient in practice. One way out consists in dealing with inferences that are weakly dependent on '$c$' when in its natural range. The range is $c = 1$–$2$ for California. One supporting remark is that '$c$' may depend on the magnitude range. Examples show that the dimension of large earthquakes is close to 1, while that of small ones is close to 2. Lastly, in scaling analysis of seismicity the magnitude $m$ and the scale $L$ are not independent, hence should be made to match.